\documentclass[aps,prl,twocolumn,showpacs,preprintnumbers,amsmath,amssymb]{revtex4}
\usepackage{dcolumn}% Align table columns on decimal point
\usepackage{bm}% bold math

\begin{document}

\title{The Mechanism of Localization in Weakly Coupled Disordered Grains}

\author{Vincent E. Sacksteder IV}
 \email{vincent@authors-last-name.com}
 \homepage[]{www.sacksteder.com}
 \affiliation{Asia Pacific Center for Theoretical Physics,  POSTECH,  Hyoja-dong, Pohang, Gyeongbuk 790-784, Korea}

\date{\today}% It is always \today, today,

\begin{abstract}
It is shown that the Wegner model of disorder contains a system of constraints which is important whenever the disorder is not weak and which is responsible for localization in $D > 2$ dimensions.  When the disorder is strong the constraints divide the phase space into an exponentially large number of domains; the localized phase is a glass.   This work is based on recently developed field theories which are exactly equivalent to the Wegner model and to an easier variant. 
\end{abstract}

%\thanks{Praise and trust to the Holy Spirit on this feast of Pentecost 2009!  In you!}

\maketitle
Disorder, meaning complicated non-repeating structure at small scales, is ubiquitous in materials of natural origin or made with imperfect human control.  Examples include sponges, clouds, cellular components, and solids containing impurities.    Weak disorder means that the fine scale structure is close to being a repeating pattern, so that the material at first examination resembles a lattice or a homogenous continuum.  Strong disorder means that no pattern or repetition can be discerned easily. Anderson pointed out that disorder can play a crucial role in regulating conduction of electrons or other quantum mechanical entities\cite{Anderson58}.   In a disordered material electrons alternate between straight unimpeded motion and bouncing off of impurities.   If they get turned around completely and recross their own path the rules of quantum mechanical interference can multiply or cancel the probability of this particular path.  As a result the electrons may be trapped, unable to move long distances, in which case the material does not conduct; this is Anderson localization.  

Weakly disordered materials do not conduct in $D = \{1,2\}$ dimensions, and do conduct otherwise\cite{GangOfFour}.  Within field theory models of disorder, this physics corresponds to spontaneous symmetry breaking; the SSB phase exhibits long distance correlations describing conduction, while the non-SSB phase has no long distance correlations\cite{Schafer80, Efetov97}.  Stronger disorder can cause a transition from conduction to localization in $D > 2$ dimensions.  The accepted supersymmetric field theory of disorder\cite{Efetov97} is mathematically unwieldy  when the disorder is not weak.  This letter analyzes stronger disorders using two recently introduced theories which  are written in terms of two $2 \times 2$ matrices, in contrast to SUSY's single $4 \times 4$ matrix. In effect half of the SUSY theory has been integrated, giving a clear mathematical picture of intermediate and strong disorders.  

\textit{The Models}--- The standard reference point for studies of conduction and localization is Wegner's model of weakly coupled grains\cite{Schafer80}.  Fyodorov showed that the zero-dimensional Wegner model is exactly equivalent to a field theory with two $2 \times 2$ matrices\cite{Fyodorov02, Fyodorov02a}.  Disertori extended Fyodorov's transformation to any dimension but used a simplified variant of the Wegner model\cite{DisertorisModel}. The Wegner model in any dimension is transformed in Ref. \cite{SackstederControlPaper}.  Both Wegner's and Disertori's models describe non-interacting electrons moving through a lattice with $V$ sites and $N$ states at each site.  The lattice geometry is completely encoded in a kinetic operator $k$ which has the $k \geq 0, \, k|\vec{0}\rangle = 0$ properties required of any Laplacian.  In Wegner's model the kinetics are constant and the potential is random;  $\langle H \rangle = \epsilon k$, $\langle (H - \langle H \rangle)^2 \rangle \propto N^{-1}$. $\epsilon$ controls the balance between kinetics and disorder.  In Disertori's model both the kinetics and the potential are random; $\langle H \rangle = 0$,  $\langle H H \rangle \propto N^{-1} (1 -k)$.  The most important difference between the two models is that Wegner's density of states $\rho$ is determined by k, while Disertori's $\rho$ is the semicircular distribution of random matrix theory.  As usual we study the two point correlator, which can be obtained by calculating the second derivative  of 
\begin{equation}{\frac{{{det}{({{\hat{E}^f_{11}}} - H)}}{{det}{({{\hat{E}^f}_{22}} - H)}}} {{det}{({{\hat{E}^b_{11}}} - H)} {det}{({{\hat{E}^b_{22}}} - H)}}}
\label{GeneratingFunction}
\end{equation} and then setting $\hat{E}^f = \hat{E}^b$.  We convert this to a path integral using $\det A \propto \int {d\bar{\psi} {d\psi}}e^{\frac{\imath}{2}\psi A \bar{\psi}}$ and ${\det}^{-1}A \propto \int {dS} {dS^*} e^{\frac{\imath}{2}S L A S^*}$, and then average over the disorder.  $L =  Sign (\, {Im} \, A)$.

  Fyodorov converted from the N-vectors  $S, \psi$ to  $2 \times 2$ matrices\cite{Fyodorov02, Fyodorov02a, Spencer04}.  The Hubbard-Stratonovich technique converts the $\psi$ vectors to an Hermitian matrix $Q^f$.  The path integral now depends on dot products $S_j^\dagger \cdot S_k$, not on individual elements of $S$. Lastly one organizes the two $S$ vectors into a $N \times 2$ matrix and performs the singular value decomposition $S = W s C$.  $C$ is $2 \times 2$ and unitary, $W$ is $ N \times 2$ and unitary, and $s$ is diagonal and positive.  These steps are exact; Disertori's model becomes\cite{DisertorisModel, SackstederControlPaper} 
\begin{eqnarray}
{\bar{Z}} & = & {\gamma_D {\int_{Q^b  \geq 0} \,{dQ^f} \,{dQ^bL} \;   e^{\mathcal{L}}  }} 
\; {\det({ Q^f  } - {  Q^b L (1 - k)   }   )} 
\nonumber \\
{\mathcal{L}}    & = &  - {N/2} \sum_{v} {Tr}\, (  Q^f_v Q^f_v +    Q^b_v L Q^b_v L)
\nonumber \\
&+& {{\imath N} \sum_v {Tr}(Q^b_v L {\hat{E}}^{b}  + Q^f_v  {{\hat{E}}^{f}} )} 
\nonumber \\
& + &   {(N - 2) \sum_v {Tr}\, \ln  Q^b_{v} L + {Tr}\, \ln  Q^f_{v} }
 \nonumber \\ 
 & - & { {N/2} \sum_{v_1 v_2} {(k/(1-k))_{v_1 v_2}}{Tr}{(Q^f_{v_1}  Q^f_{v_2}  )} } 
 \nonumber \\
& + & {{N/2}{{\sum_{v_1 v_2} {{k_{v_1 v_2}} {Tr}{(Q^b_{v_1} L Q^b_{v_2} L  )}}}}}
\label{DisertoriModel}
\end{eqnarray}
Wegner's model is exactly equivalent to\cite{SackstederControlPaper}
\begin{eqnarray}
{\bar{Z}} & = & \gamma_W \int_{Q^b  \geq 0} \,{dQ^f} \,{dQ^bL} \; {dW}  \; e^{\mathcal{L} }  \; {\det(A^0 {\delta_{j_{1}j_{2}}} - { {A^1 }  } )}
\nonumber \\
{\mathcal{L}}  & = &  - {N/2} \sum_{v} {Tr}\, (  Q^f_v Q^f_v +    Q^b_v L Q^b_v L)
\nonumber \\ 
&+&  \sum_v {{\imath N}{Tr}(  Q^f_v  {{\hat{E}}^{f}} + Q^b_v L \hat{E}^b )} 
 \nonumber \\
 & + & (N - 2) {Tr}_{vi} \,\ln (Q^f - \imath \epsilon k )
     -\imath N \epsilon  {Tr}_{vj}(\hat{S} L  k )  
\nonumber \\
&+& {(N - 2) \sum_v {Tr}\, \ln  Q^b_{v} L}
\nonumber \\
A^0 & \equiv & 
 { {Q^{f}_{v_{1} i_{1} i_{2}}}  {\delta_{v_{1} v_{2}}}}
  - {\imath \epsilon \, {\delta_{i_{1} i_{2}}}  k_{ v_1 v_2}   }, 
  \nonumber \\
 A^1 &\equiv &\sum_{i_1, v_1} {A^0_{i_0 i_1 v_0 v_1}} {(A^0)^{-1}_{i_1 i_2 v_1 v_2}}  {\hat{S}_{v_1 v_2 j_1 j_2}} {L_{j_{1}}} , \;\; 
 \nonumber \\
\hat{S}_{v_1 v_2}  & = &  C^\dagger_{v_1} s_{v_1} W^\dagger_{v_1} W_{v_2} s_{v_2} C_{v_2} , \;\;  C^\dagger_v s^2_v C_v  = Q^b_v
\label{TransformedWegnerModel}
\end{eqnarray}
 $v$ is a spatial index specifying the lattice site, while $i,j$ index the $2 \times 2$ Hermitian matrices $Q^f, Q^b$.  The argument of the $Q^f - Q^b$ determinant (on the first line of both equations) lives in a $2 \times 2 \times V$ space indexed by $ijv$.  We consider the Retarded-Advanced part of the two point correlator; $L = \sigma_3$. We decompose $Q^f = U x^f U^\dagger$ and $ Q^b L = T x^b T^{-1} $; ${dQ^f} \,{dQ^bL}  = { dU \, dx^f} \; { dT \, dx^b} \;  { \prod_v  {\Delta}_{VdM}^2(x_v^f)}  \; { {\Delta}_{VdM}^2(x_v^b)} $.  $U$ is unitary. $T$ is pseudo-unitary and unbounded, in order to preserve the sign signature of $Q^b L$'s eigenvalues\cite{Fyodorov02}.

Interactions between $U$ and $T$ are mediated only by the $Q^f-Q^b$ determinant. Leaving it aside, the $U$ sector is the classical Heisenberg model, while the $T$ sector is an hyperbolic sigma model.  Both models have been proven rigorously to exhibit SSB in $D > 2$ dimensions \cite{Frohlich76, Spencer04}.  The $T$ sector has no phase transition and always exhibits SSB\cite{Spencer04, Spencer08}.  Therefore  the determinant is responsible for all localization occuring in $D > 2$ dimensions\cite{Spencer08, SackstederControlPaper}.

\textit{Saddle point}--- A separate paper carefully applies the saddle point approximation to Disertori's model, and partially analyzes the Wegner model\cite{SackstederControlPaper}.  In summary, deviations from the saddle point are $O(N^{-1/2})$ unless $E_{11}, E_{22}$ are near the band edge or $T$ is quite large.  The latter condition is avoided if $(E_{11} - E_{22}) \pi \rho \gg 1$ and $T$ exhibits SSB.  Only in the SSB regime is the saddle point  spatially uniform.  $U, T$ are Goldstone bosons.  Disertori's saddle point is $x^f_0 = x^b_0 = L  \hat{\rho}(\bar{E}) + \imath s(\bar{E})$, where $\hat{\rho} + \imath s = e^{\imath \phi}, \sin \phi = \bar{E}/2 $.   When $\epsilon k \ll 1$ Wegner's saddle point differs by only $O(\epsilon k)$, and when $\epsilon k \gg 1$ the main difference is that $\hat{\rho}$ is $\epsilon k$'s density of states.    

\textit{Ward identity}---  By construction, the Wegner model obeys ${d\bar{Z}(\hat{E}^f = \hat{E}^b)}/{d\epsilon} = 0$.  The same Ward identity governs Disertori's model if we define $\epsilon = |k|$.   We inventory factors of $\epsilon$, assuming $\epsilon k  \ll 1$ for Wegner's but not Disertori's model.   The $W$ integral's logarithm is dominated by the self-energy $\Sigma(\vec{s}) \propto \epsilon^0 N \gg N \epsilon k(\vec{s})$.  The $x^f, x^b$ integrals, Van der Monde determinants, and remainder of the Lagrangian are controlled by $x^f_0, x^b_0 \propto 1$.  Therefore any $\epsilon$'s produced by the  $U,T$ integrals must be cancelled by the $Q^f - Q^b$ determinant.   In the SSB phase $U,T$ are controlled by the kinetic terms\cite{SackstederControlPaper}, and produce $\epsilon^{-2(V-1)}$.  In the localized phase $U,T$'s low-momentum behavior should be independent of the kinetics; the integrals should scale as $\epsilon^{-2(V-1 - \zeta V)}$, where $\zeta V$ is the volume of $U,T$'s phase space which exhibits large fluctuations. This links the determinant's behavior to localization; it must scale as $\epsilon^{2(V-1 - \zeta V)}$; the question is how? 
     
Disertori's determinant is $\det(\alpha_D = Q^f - Q^b + Q^b k)$.  In the SSB phase Wegner's determinant is $\det(\alpha_W \approx  Q^f - Q^b -\imath \epsilon k)$, plus corrections in powers of $U,T,W$'s fluctuations.  Outside of SSB one may expand $\alpha_W$ in powers of $\epsilon k \ll 1$; ignoring any resonances of $Q^f - \imath \epsilon k$ one obtains $\alpha_W \approx  Q^f - Q^b + O(\epsilon k)$.   This form can be rewritten as $\det(x^f - x^b  - \imath \epsilon k_{v_1 v_2} U_{v_1} U^\dagger_{v_2} T_{v_1} T_{v_2}^{-1})$.  $x^f$ can either cancel or add to $x^b$; consequently $x^f - x^b$ has two sectors described by two projection operators $P_l  + P_s = 1$.  Its large $P_l $ eigenvalues are $\pm 2 \hat{\rho} $, while its small $P_s$ eigenvalues are determined by $O(N^{-1/2})$ fluctuations in $x^b, x^f$ and $\hat{E}_{11} - \hat{E}_{22} \ll 1$  splitting.  If there are no angular fluctuations ($U = T = 1$)  $\alpha$ has $2(V-1)$ small eigenvalues controlled by $\epsilon k$, 2 smaller ones caused by $k|\vec{0}\rangle = 0$, and $2V$ large eigenvalues controlled by $\pm 2 \hat{\rho} + \epsilon k\,$ \cite{SackstederControlPaper}.  However because $U,T$ do fluctuate the $k U U^\dagger T T^{-1}$ term can mix the $P_l, P_s$ sectors, lifting $\zeta V$ eigenvalues to $O(2 \hat{\rho})$ to fulfill the Ward identity.  

\textit{Optimization Principle}--- The $Q^f - Q^b$ determinant is the product of $4V$ eigenvalues, each of which multiplies the entire path integral. The Van der Monde determinants  also multiply by $4V$  factors.  Therefore the path integral is governed by an optimization principle:  $Q^f$ and $Q^b$ always optimize themselves to maximize the determinants.   This principle is limited by the Lagrangian: its kinetic terms favor SSB where $U,T$'s fluctuations are $O((N \epsilon k)^{-1/2})$, while  $x^f, x^b$ fluctuations are restricted to $O(N^{-1/2})$ plus a response to $U,T$\cite{SackstederControlPaper}.  In Wegner's model when the disorder is small $\alpha_W$ is dominated by $\epsilon k$, so $4 V - 2$ eigenvalues  are pre-optimized without any need for mixing, and the kinetics ensure SSB.  The remaining $2$ very small eigenvalues caused by $k$'s  zero mode $k | \vec{0} \rangle = 0$ cause $O(N^{-1/2})$ changes in $x^f, x^b$, and are the source of the oscillatory component of Wigner-Dyson level repulsion\cite{SackstederControlPaper}. As the disorder is increased $\epsilon k$'s smallest eigenvalues become smaller than $2 \hat{\rho} \propto (\epsilon k)^{-1}$; they are no longer pre-optimized. The optimization principle starts to reward $P_s - P_l$ mixing at small momenta $\epsilon k(\vec{s}) < 2 \hat{\rho}$ and large fluctuations of $U,T$ at large wavelengths.    This trend continues into the $\epsilon k \ll 1$ regime, where the determinant favors large fluctuations at all wavelengths and the kinetic penalty on fluctuations is considerably weaker.

\textit{Constraint Structure}--- Each eigenvalue and Van der Monde factor $\lambda$ creates a forbidden hypersurface in $Q^f, Q^b$'s phase space defined by $\lambda = 0$.  Configurations on those hypersurfaces are disallowed; each hypersurface is a constraint on $Q^f, Q^b$.  A logarithmic potential $\ln \lambda$ penalises non-optimized configurations, ensuring that they do not contribute to the path integral.  In the SSB phase only two constraint hypersurfaces lie inside the accessible phase space, but as $\epsilon k$'s spectrum descends below $2 \hat{\rho}$ additional constraints are activated.  All of the $Q^f - Q^b$ constraints intersect the accessible phase space when SSB is frustrated at all wavelengths.  Leaving aside details of global topology and counting, I expect the number of disjoint domains and local minima to be exponentially large, at least $e^{4V}$ but possibly $e^{6V}$ if the Van der Monde constraints are active.  $2 (V-1)$ of the determinant's eigenvalues are $O(k \ll 1)$ when $Q^f, Q^b$ are spatially uniform, which implies a near intersection of the constraints and prohibits the existence of any spatially uniform local minimum. Fluctuations around each local minimum might be small and regulated by the kinetic terms, but the minima themselves must manifest large fluctuations.

The constraints' physical meaning is connected with the fractional form seen in equation \ref{GeneratingFunction}.  That form, including both the numerator and the denominator, was needed to properly average over all disorders, giving each possible disorder realization an equal weight.  If the disorder is small then little reweighting is necessary, but when the disorder is large reweighting becomes very important, as do the $Q^f - Q^b$ constraints.

\textit{Resonances}--- As $k$'s  lowest eigenvalues sink below $2 \hat{\rho}$, the $x^f, x^b$ eigenvalues manifest propagating resonances  which may stimulate $U,T$ to leave the SSB phase. The same resonances are manifested also in the sigma model controlling $U, T$.   In the Wegner model these resonances are manifested in $x^f, x^b$'s Hessian, $-N\delta_{v_1 v_2} - N (\imath x + \epsilon k)^{-1}_{v_1 v_2} (\imath x + \epsilon k)^{-1}_{v_2 v_1}$.   During the crossover between small and large disorder $x_0 \propto min(1, (\epsilon k)^{-1})$ passes through the spectrum of $\epsilon k$; the Hessian's kinetic component exhibits one or more resonances as $x^f_0$ passes by individual eigenvalues of $\epsilon k$.  Close to the weakly disordered regime the Hessian's kinetics are dominated by small momenta and the resonance width is small, but progressively all wavelengths become involved.   

\textit{Phases and transitions}--- We imagine starting with weak disorder and and later increasing it. Ref. \cite{SackstederControlPaper} analyzes the regime characterized by $\epsilon k_0 = E_{Th} \gg 1$ in the Wegner model and by $1 \gg k_0 \gg N^{-1/2}$ in Disertori's model, where $k_0$ is $k$'s smallest non-zero eigenvalue.  In this regime mathematical control over Disertori's model has been obtained, and a partial analysis suggests that the Wegner model might also be controlled.  The results reproduce and extend standard SUSY results concerning the two point correlator, anomalously localized states, and localization in $D = \{1, 2\}$ dimensions.  The $Q^f - Q^b$ determinant is nearly static,  but its vestiges are responsible for Wegner-Dyson statistics.  The dominant physics is scattering, and is encoded in the sigma model Lagrangian controlling $U, T$.  The Mermin-Wagner theorem\cite{Mermin66} says that scattering ensures localization in large volumes in $D = \{1, 2\}$ dimensions. The scattering by itself is unable to achieve localization in $D > 2$ dimensions, but is able to produce individual anomalously localized states.    

As the disorder is increased $2 \hat{\rho}$ approaches $\epsilon k_0$ and $x^f, x^b$ develop critical fluctuations at long wavelengths.  Crossing this threshold activates the first of many new constraints, each of which pushes $U, T$  to develop large long-wavelength fluctuations with the help of $x^f, x^b$'s resonance. 

A straightforward analysis suggests that the new constraints become active in sequence starting with long wavelengths, and that the mobility edge is a function of the wavelength, with conduction at small scales and localization at large scales.  However this is not the last word: the model might be able to maximize a few  long-wavelength constraints without making a full transition from $O((N\epsilon k)^{-1/2})$ to $O(1)$ fluctuations; in this case one might see subdiffusive conduction below some threshold number of constraints and localization above that threshold.  Alternatively, some collective dynamics might make the model unresponsive to new constraints until a threshold is reached; constraint optimization could occur en masse in a single Anderson transition.  What is certain is that at large enough disorder SSB is completely frustrated with large fluctuations at all wavelengths and all of the $Q^f - Q^b$ determinant's constraints are active. 

\textit{Renormalization group}---  $k$'s smallest eigenvalues scale as $V^{-2/D}$; therefore increasing the length scale is roughly equivalent to increasing the disorder.  The renormalization group starts in the weak disorder phase and sees resonances and new constraints as the volume is increased.  However $k$'s large eigenvalues are independent of $V$, so the $V \rightarrow \infty$ limit brings only a fraction of $k$'s eigenvalues below $2 \hat{\rho}$.  Anderson localization vs. diffusive or subdiffusive conduction is determined by whether that fraction is able to frustrate SSB at long wavelengths. 

We compare this picture to the SUSY picture of renormalization\cite{Efetov83}.  In the SUSY approach the renormalization group equation's initial value is the SUSY sigma model, which is a weak-disorder approximation of the Wegner model.  The standard SUSY sigma model results (without renormalization) can be reproduced by approximating the $Q^f - Q^b$ determinant to the point where only two constraints are active\cite{SackstederControlPaper}.  It would be remarkable if the SUSY renormalization group could recover the physics of the full $Q^f - Q^b$ determinant from this very simplified starting point. 
  
The renormalization group may be at least partly superfluous, since the models considered here are exactly equivalent to the original Wegner and Disertori models including all band and geometric information. They avoid the Grassman degrees of freedom found in the SUSY model, but by exactness retain their physics; in this sense the $Q^f - Q^b$ determinant should be duplicated, not improved upon, by an exact integration of the SUSY renormalization group.

\textit{Disertori's Model}---  This model gives some hints of an Anderson transition around the  $k_0 \approx N^{-1/2}$ threshold.  At this point the spectral gap beween $k$ and  $x^f - x^b$'s small eigenvalues disappears, which facilitates mixing, although the mixing's effects may be limited by $N^{-1/2}$.  
More interestingly, at the threshold $x^f, x^b$ undergo a phase transition as they begin to optimize the $Q^f - Q^b$ determinant's eigenvalues. The order parameter is an $O(N^{-1/2})$ splitting between  $x^f$ and $x^b$ enforced by the constraints' logarithmic repulsion.  A naive computation of the eigenvalues' Hessian using this splitting reveals a pole on the energy shell $k(\vec{s}) \approx N^{-1/2}$; $x^f, x^b$ organize themselves across long distances in order to set an $O(N^{-1/2})$ lower bound  on the determinant's eigenvalues. The resonance's length scale becomes progressively smaller as $k$ decreases. Its width is $O(N^{-1/2})$, since its interaction with $U,T$ is controlled by derivatives of the determinant's logarithm, which are proportional to inverse powers of the splitting.    

As mentioned earlier there are exponentially many local minima corresponding to opposite signs of the splitting; determining the resonance's effects on $U,T$ requires first integrating $x^f, x^b$.  Here lies a mystery. If SSB controls $T$'s unboundedness then the saddle point integral is Gaussian.   It is a simple matter to establish that the averaged determinant is bounded below by $N^{-2V}$, not by the $N^{-V}$ factor one would expect from the eigenvalue splitting; on average the splitting is invisible! Adding to the mystery, this disappearing act occurs only in calculations of two-point correlators and not in higher order correlators, i.e. if one starts with three or more factors in equation \ref{GeneratingFunction} then the averaged determinant clearly manifests the eigenvalue splitting. 

If the Anderson transition does not occur at $k_0 \approx N^{-1/2}$, it must occur at $k \approx N^{-1}$, because after this point the kinetic term $Nk$ is unable to regulate fluctuations.   If one pushes further to the point where $k$'s spectrum is partly or completely negative, the kinetic term rewards $U,T$'s fluctations rather than penalizing them.  

\begin{acknowledgments}

Giorgio Parisi and Tom Spencer have been very influential.  Discussions with Margherita Disertori, Yan Fyodorov, Martin Zirnbauer, Jacobus Verbaarschot, M.A. Skvortsov, Alexander Mirlin, and Konstantin Efetov were very helpful.  I am grateful for support from the Isaac Newton Institute, the Universit\`a degli Studi di Roma "La Sapienza", the ICTP, and the APCTP.  Most of this work was performed at the APCTP.

\end{acknowledgments}

\bibliography{vincent}

\end{document}